\documentclass[pdflatex,sn-mathphys]{sn-jnl}% Math and Physical Sciences Reference Style
\jyear{2022}%

\theoremstyle{thmstyleone}%
\theoremstyle{thmstyletwo}%

\theoremstyle{thmstylethree}%

\usepackage{color}

\usepackage{float}
\usepackage{etoolbox}
\makeatletter
\patchcmd{\ps@headings}
{\hbox to \hsize{\hfill Springer Nature 2021 \LaTeX\ template\hfill}}
{\hbox to \hsize{}}
{}
{}
\patchcmd{\ps@headings}
{\hbox to \hsize{\hfill Springer Nature 2021 \LaTeX\ template\hfill}}
{\hbox to \hsize{}}
{}
{}
\patchcmd{\ps@titlepage}
{\hbox to \hsize{\hfill Springer Nature 2021 \LaTeX\ template\hfill}}
{\hbox to \hsize{}}
{}
{}
\makeatother

\begin{document}
\small

\title[ ]{\Large Super-resolution imaging using super-oscillatory diffractive neural networks}


\author[1]{\normalsize \fnm{Hang} \sur{Chen}}
\equalcont{\normalsize These authors contributed equally to this work.}

\author[1]{\normalsize \fnm{Sheng} \sur{Gao}}
\equalcont{\normalsize These authors contributed equally to this work.}

\author[1]{\normalsize \fnm{Zejia} \sur{Zhao}}
% \equalcont{\normalsize These authors contributed equally to this work.}

\author[1]{\normalsize \fnm{Zhengyang} \sur{Duan}}

\author[1]{\normalsize \fnm{Haiou} \sur{Zhang}}

\author[2]{\normalsize \fnm{Gordon} \sur{Wetzstein}}

\author*[1,3]{\normalsize \fnm{Xing} \sur{Lin}}

\affil[1]{\orgdiv{\normalsize Department of Electronic Engineering}, \orgname{Tsinghua University}, \orgaddress{\city{Beijing}, \postcode{100084}, \country{China}}}

\affil[2]{\orgdiv{\normalsize Department of Electrical Engineering}, \orgname{Stanford University}, \orgaddress{\city{California}, \postcode{94305}, \country{USA}}}

\affil[3]{\orgdiv{\normalsize Beijing National Research Center for Information Science and Technology}, \orgname{Tsinghua University}, \orgaddress{\city{Beijing}, \postcode{100084}, \country{China}}}

%\affil[2]{\orgdiv{\normalsize Beijing National Research Center for Information Science and Technology}, \orgname{Tsinghua University}, \orgaddress{\city{Beijing}, \postcode{100084}, \country{China}}}

\email{\normalsize lin-x@tsinghua.edu.cn}

%%================================%%
%% Abstract %%
%%================================%%

\abstract{Optical super-oscillation enables far-field super-resolution imaging beyond diffraction limits. However, the existing super-oscillatory lens for the spatial super-resolution imaging system still confronts critical limitations in performance due to the lack of a more advanced design method and the limited design degree of freedom. Here, we propose an optical super-oscillatory diffractive neural network, i.e., SODNN, that can achieve super-resolved spatial resolution for imaging beyond the diffraction limit with superior performance over existing methods. SODNN is constructed by utilizing diffractive layers to implement optical interconnections and imaging samples or biological sensors to implement nonlinearity, which modulates the incident optical field to create optical super-oscillation effects in 3D space and generate the super-resolved focal spots. By optimizing diffractive layers with 3D optical field constraints under an incident wavelength size of $\lambda$, we achieved a super-oscillatory spot with a full width at half maximum of 0.407$\lambda$ in the far field distance over 400$\lambda$ without side-lobes over the field of view, having a long depth of field over 10$\lambda$. Furthermore, the SODNN implements a multi-wavelength and multi-focus spot array that effectively avoids chromatic aberrations. Our research work will inspire the development of intelligent optical instruments to facilitate the applications of imaging, sensing, perception, etc.}

%, e.g., 3$\times$5 spot array, over red, blue, and green channels
% Our innovation reforms the complex inverse design for optical super-oscillation into the process of training large-scale diffractive neural networks with the super-resolved resolution, which
% An optical super-oscillatory needle with a  demonstrates that SODNN with a high degree of freedom can achieve more complex optical super-oscillation designs in 3D




\maketitle

\section*{1 Introduction}\label{Main}

The Abbe-Rayleigh diffraction limit of traditional optical equipment has always been an obstacle to the study of micro-/nano-scale objects \cite{s1, s2}. Near-field microscopic imaging techniques, such as contact photography \cite{r1} and scanning near-field imaging (SNOM) \cite{r2,s3}, capture evanescent fields by placing a probe or light-sensitive material extremely close to the object to achieve nanoscale resolution, which is not possible for imaging inside biological samples or encapsulated micro-/nano-structures. Far-field microscopic imaging technology is not restricted by the above bottlenecks. Some typical far-field microscopic imaging techniques, such as single-molecule localization (SML) microscopy \cite{r3, s4} or stimulated emission depletion (STED) \cite{r4, s5}, have demonstrated the possibility of nanoscale imaging without capturing evanescent fields. However, SML microscopy and STED typically require intense beams to excite, deplete, or bleach fluorophores in a sample that produces photobleaching and phototoxicity in living samples.

Optical super-oscillations are rapid sub-wavelength spatial variations of light intensity and phase that occur in complex electromagnetic fields formed by the precise interference of coherent waves, which provide an advanced method for far-field super-resolution imaging beyond the diffraction limit \cite{r6, r7}. To generate optical super-oscillation, the complicated lens design methods \cite{q1, q2, q7} or Fresnel zone plate (FZP) optimization design methods, including optimizing algorithms \cite{r5, q3, q5, b1} or optimization-free algorithms \cite{q4, r8}, have been proposed. However, the existing super-oscillatory focusing-based imaging devices still have limited performance as follows: (1) strong side-lobes resulting in a small field of view (FoV) \cite{r5, r8, q1}, (2) short working distances \cite{r5, q2, q3, q4}, (3) limited depth-of-focus (DoF) \cite{q5, r5, q6, b1}, and (4) chromatic aberration caused by wavelength-dependent phase retardation \cite{q2, q7, q12, q4}, which dramatically limit their applications.

\begin{figure*}[!t]
    \centering
    \includegraphics[width=1.0\textwidth]{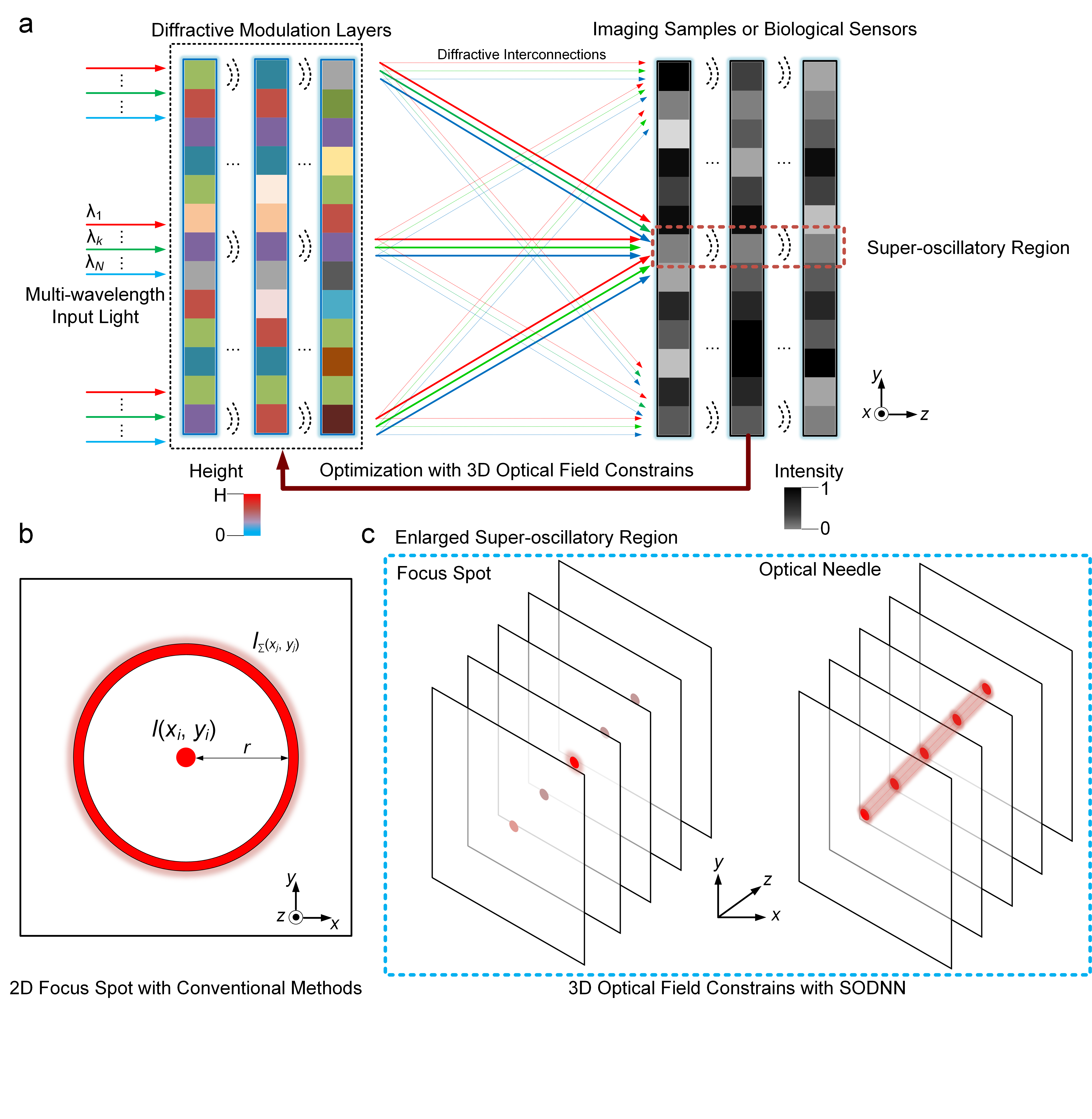}
    \caption{\textbf{Training SODNN to optimize the diffractive elements with 3D optical field constrains.} (a) Utilizing diffractive modulation layers and free-space propagation to implement the weighted optical interconnections and imaging samples or biological sensors to implement nonlinearity. SODNN can modulate the multi-wavelength incident optical field to create optical super-oscillation effects in 3D super-oscillatory regions. (b) The conventional methods optimize a 2D focus spot at a specific focusing distance to achieve optical super-oscillation with a large side lobe. (c) The enlarged 3D super-oscillatory regions show that SODNN optimizes the 3D optical field in a certain distance range to achieve super-oscillation without the side lobe.}
    \label{figure_1}
\end{figure*}

The performance limitations of the existing super-oscillatory imaging methods are mainly due to the optimization with 2D optical field constraints with limited modulation element numbers or only ring-structured phase modulation, which substantially limits the degree of freedom in design space for optimizing the performance. For example, the conventional super-oscillatory lens design methods, such as the pinhole array mask \cite{r7} or FZPs \cite{q5}, require the following steps. The prolate ellipsoid function or Strehl ratio is first used as the optimization function of the model. Then, the phase distribution of the super-oscillatory lens is optimized with the constraints of full width at half maximum (FWHM) of the super-oscillatory spot $I_{(x_\mathrm{i}, y_\mathrm{i})}$ at the 2D position of $(x_\mathrm{i}, y_\mathrm{i})$, and the side-lobes intensity $I_{\sum_{(x_\mathrm{j}, y_\mathrm{j})}}$ at a combination of the 2D positions $\sum_{(x_\mathrm{j}, y_\mathrm{j})}$ in the local FoV expressed by the distance \emph{r} between the super-oscillatory spot and side-lobes, as shown in Fig.~\ref{figure_1}(b). The above optimization method can only achieve the design of a pinhole array mask with a limited element number or a simple FZP with the ring-structured phase modulation of 0 or 1, while the optimization process requires a very complex formula decomposition process.

Here, we propose constructing the super-oscillatory diffraction neural networks, i.e., SODNN, that generate optical super-oscillation in 3D and achieve super-resolution imaging beyond diffraction limits. SODNN is constructed by utilizing diffractive layers to implement optical interconnections and imaging samples or biological sensors to implement nonlinearity, as shown in Fig.~\ref{figure_1}(a). SODNN modulates the incident optical field to create optical super-oscillation effects in 3D space and generates the super-resolved focusing spots or optical super-oscillatory needle, as shown in Fig.~\ref{figure_1}(c). The diffraction limit of existing photonic neural network systems is due to training neural networks without exploring the super-oscillatory effects. By constructing the large-scale SODNN that optimizes the optical coefficients of a stack of diffractive layers to modulate the optical field in 3D space, we can generate super-oscillation at any local regions without side-lobes across the FoV and with long working distance, long DoF, and achromatic spots for high-performance super-resolution imaging.

%\subsection{Optimization of SODNN beyond the diffraction limit}
%\label{Optimization of SODNN}

\section*{2 Methods}\label{methods}

%\textbf{2.1 Formulation of SODNN training steps}

The forward model of SODNN is based on the angular spectrum representation. The complex-valued coherent input optical field $U_{{\lambda_k}}$ at the wavelength ${\lambda_k}$ $(\textit{k}=1, 2, ... , \textit{N})$ are transformed by SODNN before the detection. We consider the diffractive modulation layers of SODNN in this work with the complex transform function of ${M_{{\lambda}_k}}{(\Delta\emph{H}, z_i)}$ that modulates the incident optical field to the output optical field at the plane with a distance of $z_i$. Here, $\Delta \emph{H}$ represents the relative height map of the diffractive elements in SODNN to generate optical path difference and modulate the phase $\phi_{{\lambda}_k}$ of incident optical field, which can be formulated as $\Delta \emph{H} = {{\lambda}_k}{\bf{\phi}_\lambda}_k / 2 \pi \Delta \emph{n}_{{\lambda}_k}$ with $\Delta\emph{n}_{{\lambda}_k}$ being the wavelength-depended material refractive index. Then, the output optical fields at the wavelength ${\lambda}_k$ at the output plane with a distance of $z_i$ can be formulated as ${U}_{{\lambda}_k}^{'}(z_i) = {M}_{{\lambda}_k}(\Delta\emph{H}, z_i){U}_{{\lambda}_k}$, and the detector measures the intensity distribution of output optical fields that can be formulated as a nonlinear function such as the square of the complex optical field: ${I}_{{\lambda}_k}(z_i) = \vert {U}_{{\lambda}_k}^{'} (z_i) \vert ^2 = \vert {M}_{{\lambda}_k}(\Delta\emph{H}, z_i){U}_{\bf{\lambda}_k} \vert ^2$. As the relative height map of each diffractive modulation layer is the same under different wavelength channels, the above model can effectively eliminate chromatic aberration caused by wavelength-dependent phase retardation \cite{s7}. For the multi-wavelength SODNN, the total intensity distribution of different wavelengths at the output plane can be formulated as the superposition of detected intensity distribution at each wavelength: ${I}(z_i) = \sum_{{\lambda}_k}{I}_{{\lambda}_k}(z_i)$.

The SODNN is optimized with the 3D optical field constrain, which optimizes the shape of the super-oscillatory focusing spot within a certain distance range $z_{i}$ $\in$ [\textit{f}-$\Delta$\textit{f}, \textit{f}+ $\Delta$\textit{f}] before and after the focal plane with a focal length of \textit{f}. The ideal super-oscillatory focusing spot $I_{(x_{i}, y_{i}, z_{i})}$  at a position in the 3D space with coordinates $(x_{i}, y_{i}, z_{i})$ at the output plane with a distance of $z_{i}$ would have a maximized energy focused spot $I_{(x_{i}, y_{i}, z_{i})}$ and minimized light intensity side-lobes $I_{\sum_{\substack{j}} {(x_{j}, y_{j}, z_{i})}}$ at a combination of position coordinates $\sum_{\substack{j}} {(x_{j}, y_{j}, z_{i})}$. Taking the ideal output light intensity as the optimization direction with the entire 3D optical field constraints, the SODNN performs the function of a neuromorphic photonic processor that utilizes weighted optical diffractive interconnections of massively diffractive neurons to achieve the desired optical super-oscillatory function. Besides, we further design the constraint to maximize the energy transmission efficiency of super-oscillatory regions by minimizing the optical energy outside the super-oscillatory regions. Therefore, the 3D optical field constrain optimization of SODNN can be formulated as:
\begin{equation}\label{mse_loss}\tag{1}
    \min_{\Delta\emph{H}} (\sum_{\substack{z_{i}\in [f-\Delta f, f+ \Delta f]}} {\left\{((I_{{(x_{i}, y_{i}, z_{i})}} + I_{\sum_{\substack{j}}{(x_{j}, y_{j}, z_{i})}})-I_{target})^2 + MSE(I_{(x, y, z) \notin {(x_{i}, y_{i}, z_{i})}}) \right\}}),
\end{equation}
where $I_{(x_{i}, y_{i}, z_{i})}$ is the intensity of the super-oscillatory focusing spot at the 3D position of $(x_{i}, y_{i}, z_{i})$; $I_{\sum_{\substack{j}} {(x_{j}, y_{j}, z_{i})}}$ is the intensity of the side-lobes at the 3D positions of $\sum_{\substack{j}} {(x_{j}, y_{j}, z_{i})}$; $I_{target}$ is the ground truth label which represents the ideal super-oscillatory output; $MSE(I_{(x, y, z) \neq (x_{i}, y_{i}, z_{i})})$ represents the total energy of optical intensity outside the super-oscillatory regions with the mean square error (MSE) function;  $z_{i}$ $\in$ [\textit{f}-$\Delta$\textit{f}, \textit{f}+$\Delta$\textit{f}] is the range of the 3D optical field constrain optimization space before and after the focal plane with a focal length of \textit{f}; and $\Delta$\textit{H} is the relative height map of the diffractive elements.

For the design of SODNN in this work, we use the stochastic gradient descent approach to optimize the network coefficients on a desktop computer (Linux) with an Intel Xeon Gold 6226R CPU at 2.90GHz with 16 cores and an Nvidia GTX-3090Ti GPU of 24 GB graphics card memory. The residual error of network outputs with respect to ground truth labels and the total optical energy outside the super-oscillatory regions are calculated according to Eq.~\eqref{mse_loss}, which are used to perform the error back-propagation to optimize the SODNN and the relative height map $\Delta$\textit{H} of the diffractive modulation elements in SODNN.

\section*{3 Results}\label{results}

\subsection*{3.1 Numerical Evaluations}\label{3_1}

\begin{figure*}[!h]
    \centering
    \includegraphics[width=1.0\textwidth]{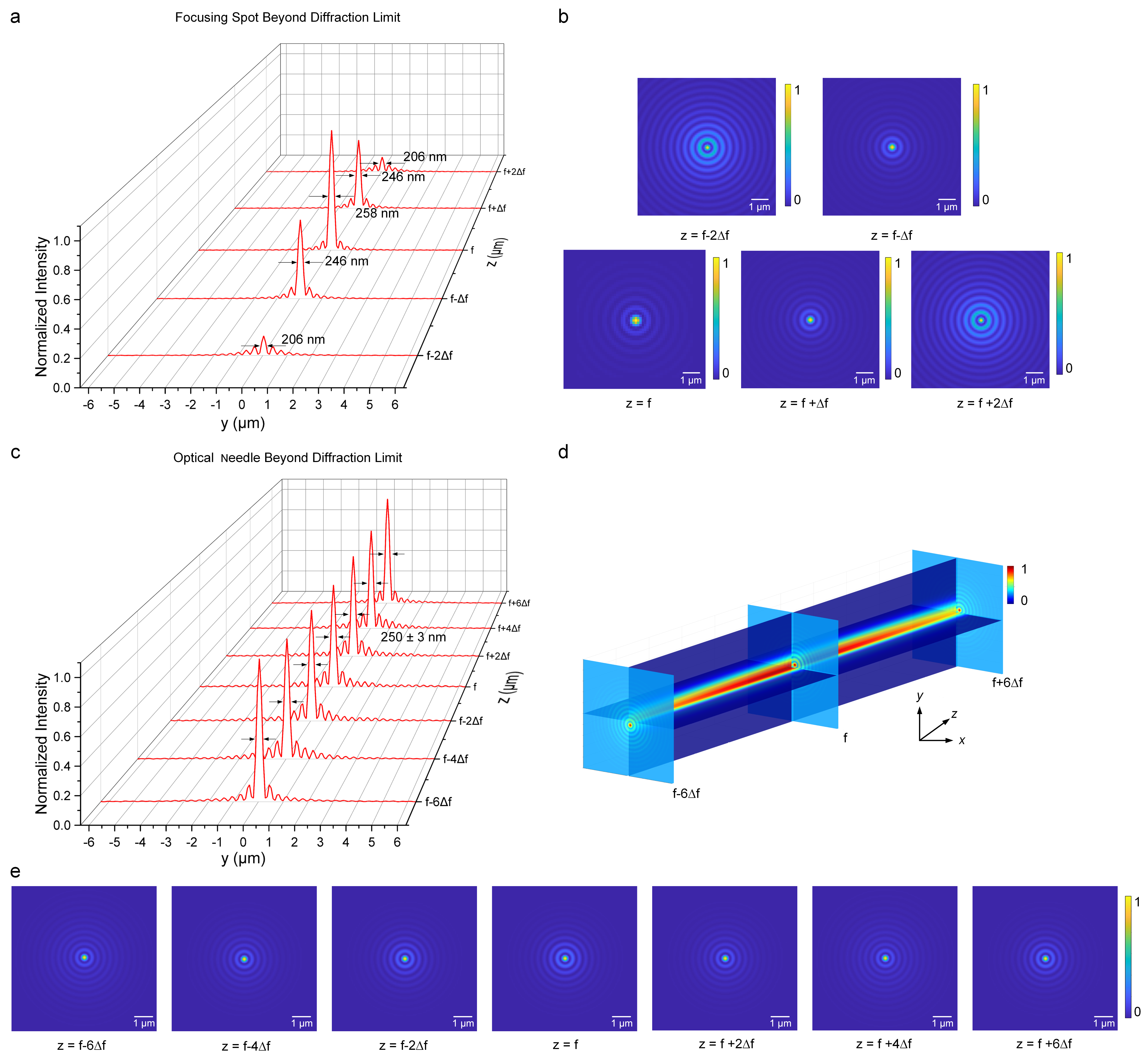}
    \caption{\textbf{Optical super-oscillatory spots and optical needle design of SODNN}. The FWHM (a) and the output (b) of super-oscillatory spots at the designed focal length and distributions offsetting the designed focal length with the collimated input optical field. (c) The optical super-oscillatory needle within a DoF of 6 $\mu$m with uniform light intensity and consistent FWHM. (d) The 3D distributions of the output optical super-oscillatory needle. (e) The output of the slices of the optical super-oscillatory needle.}
    \label{figure_2}
\end{figure*}

We first validate the effectiveness of the 3D optical field constrain optimization of SODNN in achieving super-oscillatory spots without side-lobes to realize large FoV at the designed long focal length \emph{f}, as shown in Fig.~\ref{figure_2}(a) and Fig.~\ref{figure_2}(b). Each optical diffractive element size was set to $\lambda$/2 × $\lambda$/2 where $\lambda$ is the wavelength of input coherent light and $\lambda$=632.8 $nm$. We designed a 1-layer SODNN by setting the modulation element number to 2500 × 2500, corresponding to the network layer size of ~0.79 $mm$ × 0.79 $mm$. The selection of the above parameters is determined according to the performance analysis of SODNN, as discussed in Section 4.2. The proposed system forms a super-oscillatory focused spot with almost no side-lobes at a long focal length $\emph{f}=250 \; \mu m (\sim400\lambda$), with a FWHM of 258 $nm$ ($\sim0.407\lambda$, see the middle of Fig.~\ref{figure_2}(a) and the lower left of Fig.~\ref{figure_2}(b)). Recent achievements of the super-oscillatory lenses compared with SODNN are shown in Section 4.1, which demonstrates the extraordinary performance of SODNN.

It can be found that optimizing the SODNN only at a designed focal length \emph{f} (that is $\Delta\textit{f}=0$ in Eq.~\eqref{mse_loss}), SODNN can maintain the morphology of the super-oscillatory focus spot and avoid the appearance of side-lobes at the designed position, once it exceeds this distance, the side-lobes will appear immediately accompanied by smaller FWHM and weaker super-oscillatory focusing spot intensity. For example, at four different focal lengths $\emph{f}-2\Delta \emph{f}$, $\emph{f}-\Delta \emph{f}$, $\emph{f}+\Delta \emph{f}$, and $\emph{f}+2\Delta \emph{f}$ where $\Delta \emph{f}=0.25 \; \mu m$, we found that the side-lobes appeared and the FWHM was reduced to 206 $nm$($\sim0.325\lambda$), 246 $nm$($\sim0.388\lambda$), 246 $nm$($\sim0.388\lambda$), and 206 $nm$($\sim0.325\lambda$), respectively, as shown in Fig.~\ref{figure_2}(a) and Fig.~\ref{figure_2}(b). For the out-of-focal planes, the intensity of the side lobes increases exponentially with respect to the central spot as the central spot size decreases, demonstrating the high quality of optical sectioning of the SODNN for imaging.

In order to maintain the profiles of super-oscillatory light spots within a long DoF range, that is, to form a super-oscillatory light needle, we optimize the shape of the super-oscillatory focusing spot within a certain distance range $[\emph{f}-6\Delta \emph{f}, \emph{f}+6 \Delta \emph{f}]$ before and after the focal length \textit{f} as shown in Fig.~\ref{figure_2}(c), Fig.~\ref{figure_2}(d) and Fig.~\ref{figure_2}(e). Each optical diffractive element size was set to $\lambda$/2 × $\lambda$/2 where  $\lambda=632.8 \; nm$. We evaluate the performance of 1-layer SODNN by setting the modulation element number to 1500 × 1500, corresponding to the network layer size of ~0.47 $mm$ × 0.47 $mm$. We set the range of the 3D optical field constrain optimization space $z_\mathrm{i} \in [\emph{f}-6\Delta\emph{f}, \emph{f}+6\Delta\emph{f}]$ where $\emph{f} = 100 \; \mu m$ and $\Delta\emph{f}=0.5 \; \mu m$. We found that an optical super-oscillatory needle was formed within a long DoF of 6 $\mu$m ($\sim10\lambda$, see Fig.~\ref{figure_2}(d)). The optical super-oscillatory needle was further tested by selecting 7 positions in the range $ [\emph{f}-6 \Delta \emph{f}, \emph{f}+6 \Delta \emph{f}]$ with 2$\Delta$\emph{f} as the sampling interval (see the slices of the optical super-oscillatory needle in Fig.~\ref{figure_2}(e) and the FWHM in Fig.~\ref{figure_2}(c)). It can be found the optical super-oscillatory needle has uniform light intensity and consistent FWHM (250 $nm$ ± 3 $nm$) in the designed DoF.

\begin{figure*}[!h]
    \centering
    \includegraphics[width=1.0\textwidth]{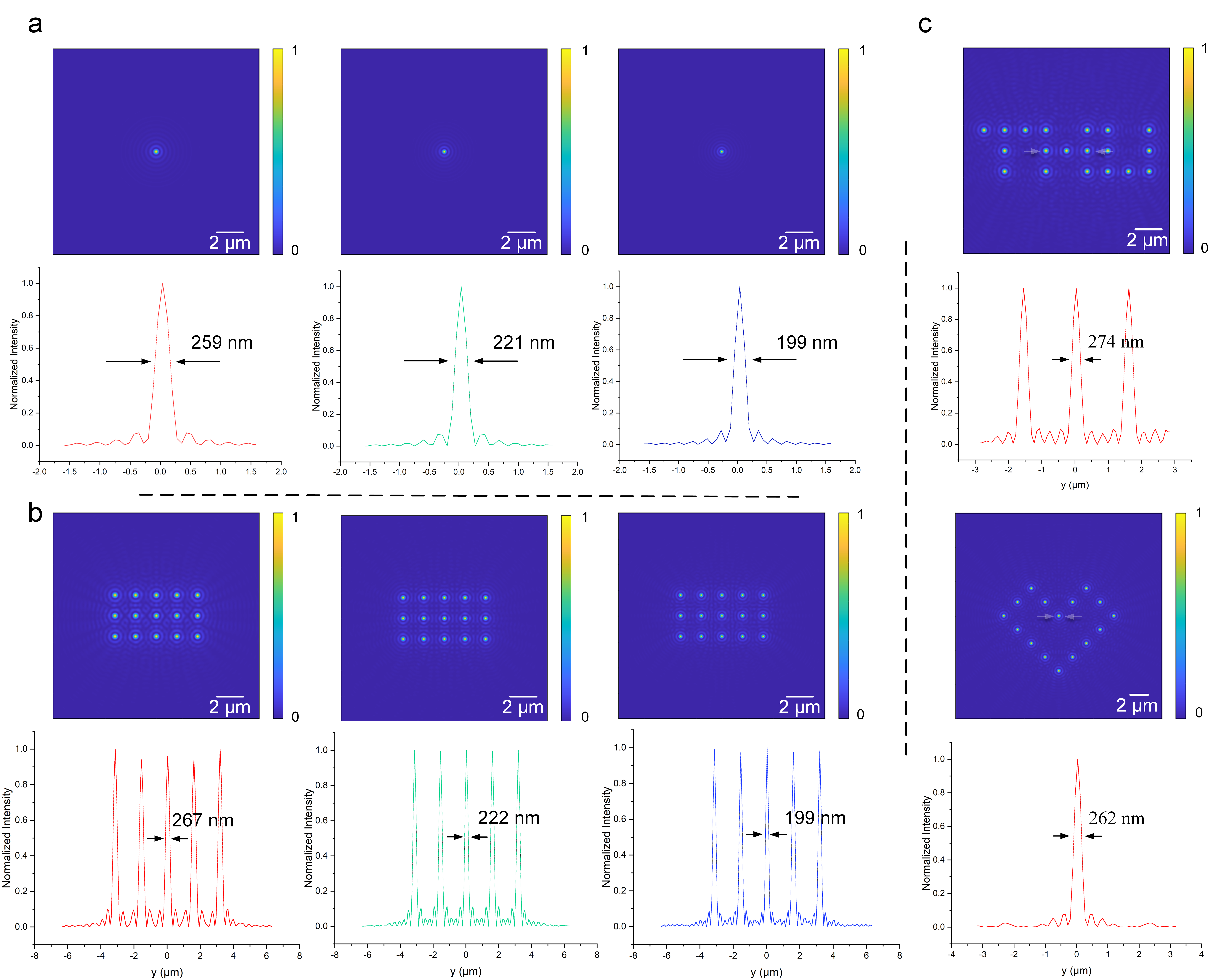}
    \caption{\textbf{Multi-wavelength and multi-focus SODNNs}. (a) The super-oscillatory spots under red, green, and blue light channels with the FWHM of 259 $nm$, 221 $nm$, and 199 $nm$, respectively. (b) The 3 × 5 super-oscillatory spot arrays under red, green, and blue light channels with the FWHM of 267 $nm$, 222 $nm$, and 199 $nm$, respectively. (c) The super-oscillatory spots of the T-H-U pattern with the FWHM of 274 $nm$ and the super-oscillatory spots of the heart-shaped pattern with the FWHM of 262 $nm$.}
    \label{figure_3}
\end{figure*}

Grounded on our earlier studies \cite{s7, chen2021}, we further demonstrate the multi-wavelength SODNN to solve the chromatic aberration problems caused by wavelength-dependent phase retardation as shown in Fig.~\ref{figure_3}(a) and Fig.~\ref{figure_3}(b). In this design, 3 different parallel wavelength channels (e.g. blue, green and red light with the wavelengths of 473 $nm$ ($\lambda_1$), 532 $nm$ ($\lambda_2$), and 632.8 $nm$ ($\lambda_3$) respectively) are used to generate multi-wavelength super-oscillatory light spots focused at the same focal length. Each optical diffractive element size was set to (${\lambda}_3)/2 \times ({\lambda}_3)/2$. We also designed a 1-layer SODNN by setting the modulation element number to 2500 × 2500, corresponding to the network layer size of ~0.79 $mm$ × 0.79 $mm$. The proposed system forms multi-wavelength super-oscillatory focused spots with almost no side-lobes at a long focal length $\emph{f}=250 \; \mu m (\sim400 \lambda_3)$, with the FWHM of 259 $nm$, 221 $nm$, and 199 $nm$ respectively produced by red light, green light and blue light shown in Fig.~\ref{figure_3}(a) respectively. Integrated multi-focus and multi-wavelength design approach proposed above, we further design a multi-wavelength multi-focus SODNN shown in Fig.~\ref{figure_3}(b), which can realize 3 × 5 super-oscillatory focusing spot arrays under red, green, and blue light channels. The 3 × 5 super-oscillatory focusing spot arrays were produced by red light, green light, and blue light with the FWHM of 267 $nm$, 222 $nm$, and 199 $nm$, respectively. State-of-the-art achievements of super-oscillatory design compared with the proposed SODNN are also discussed in the Discussion section 4.1. Some complex multi-focus super-oscillatory spots, such as heart-shaped pattern and  T-H-U  pattern (the abbreviation of  Tsinghua University), are shown in Fig.~\ref{figure_3}(c) to demonstrate the superiority of SODNN in terms of design flexibility and versatility. For the above two multi-focus arrays, we obtained the FWHM of 274 $nm$ and 262 $nm$, respectively, beyond the diffraction limit (0.61$\lambda$/NA=456 $nm$).

%\subsection{Characterizing SODNN performance}
%\label{Characterization_SODNN}

\begin{figure*}[!t]
    \centering
    \includegraphics[width=0.8\textwidth]{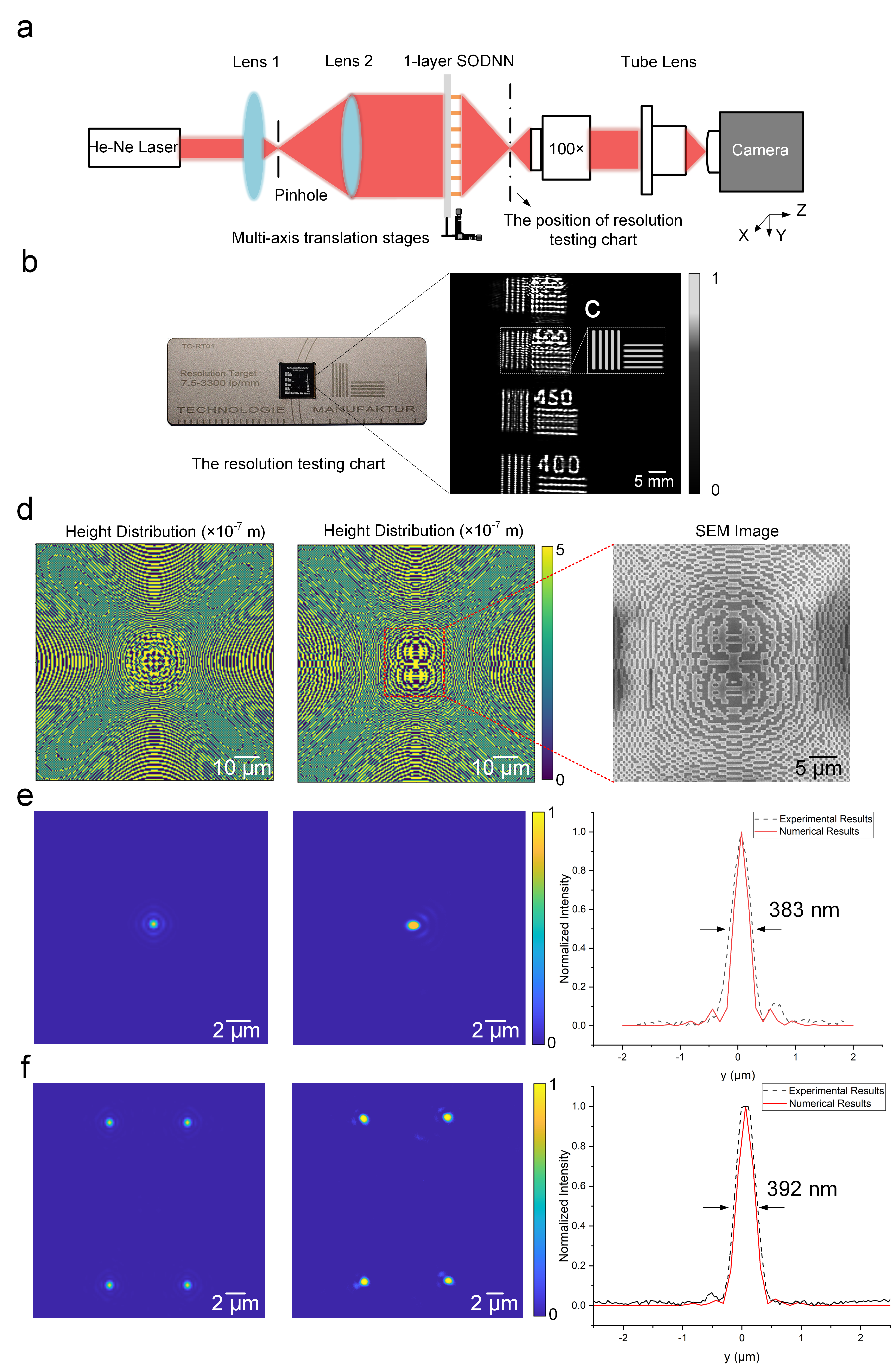}
    \caption{\textbf{Characterization of SODNN.} (a) Schematic of the experimental setup. (b, c) Imaging results of the resolution testing chart by commercial Olympus objective and SODNN. (d) The diffractive modulation layer of single-layer SODNNs for a single-focus (left) and 2 × 2 multi-focus (middle) with the layer profile characterized by scanning electron microscope, i.e., SEM (right). (e) The numerical analysis results (left) and experimental results (middle) of the single-focus SODNN. (f) The numerical analysis results (left) and experimental results (middle) of the 2 × 2 multi-focus SODNN.}
    \label{figure_4}
\end{figure*}

\subsection*{3.2 Experimental Evaluations}\label{3_2}

We built the experimental setup to measure the SODNN’s super-oscillatory focusing spot profiles, as shown in Fig.~\ref{figure_4}(a). A He–Ne laser (25-STP-912-230, Melles Griot, USA) with 632.8 $nm$ wavelength and 5 $mW$ power was collimated by Lens 1 and Lens 2, and a pinhole was used as a filter. The collimated He–Ne laser beam was used to illuminate the SODNN placed on the multi-axis translation stage (NPXYZ100SGV6, Newport, USA), which was processed by two-photon printing technology (MOJI-NANO TECHNOLOGY CO, China). Considering the error accumulation and processing costs of large-scale devices in actual processing, we designed two kinds of small-scale 1-layer SODNN with the modulation element number of 200 × 200 corresponding to each optical diffractive element size of 500 $nm$ × 500 $nm$ and 1-bit step height of 0 $nm$ and 500 $nm$ to verify the correctness of SODNN. Fig.~\ref{figure_4}(d) left and middle are respectively designed to achieve single-focus focusing and 2 × 2 multi-focus array focusing of super-oscillatory spots while Fig.~\ref{figure_4}(d) right is the enlarged part of the processed 2 × 2 multi-focus SODNN after characterization by Electron Microscopy (EM). An Olympus objective (100× magnification, NA=0.9) was used to image the light focused by the designed SODNNs. A tube lens with focal length \emph{f} = 180 $mm$ was used to form an image on a CMOS camera (01-MOMENT, Photometrics, USA).

Fig.~\ref{figure_4}(e) left and Fig.~\ref{figure_4}(f) left are the numerical analysis results of a single-focus super-oscillatory focusing spot and 2 × 2 multi-focus super-oscillatory focusing spot array with a focal length $\emph{f}=20 \; \mu m$ at its design wavelength $\lambda$ = 632.8 $nm$. The spot morphology is highly consistent with the experimental measurement results shown in Fig.~\ref{figure_4}(e) middle and Fig.~\ref{figure_4}(f) middle. For the single-focus super-oscillatory focusing spot and the 2 × 2 multi-focus super-oscillatory focusing spot array, the FWHMs obtained by numerical analysis results are 307 $nm$ (see Fig.~\ref{figure_4}(e) right) and 340 $nm$ (see Fig.~\ref{figure_4}(f) right), respectively. However, due to the systematic errors during the optical processing, the experimentally measured FWHMs were 383 $nm$ (see Fig.~\ref{figure_4}(e) right) and 392 $nm$ (see Fig.~\ref{figure_4}(f) right), respectively. Although there was a slight gap with the numerical results, the experimental results still exceeded the diffraction limit (0.61$\lambda$/NA=415 $nm$).

For imaging, we used a scanning mode with SODNN, where the signal used to reconstruct the image is taken from the central part of the CMOS camera. Such imaging strategy is also employed in confocal microscopy \cite{r15, r16}. The experimental equipment is shown in Fig.~\ref{figure_4}(a), where the commercial resolution testing chart (TC-RT01, Technology Manufacture, Germany, see Fig.~\ref{figure_4}(b)) is used as the measured targets. The resolution testing chart's pattern structure is achieved by processing metallic chromium (Cr) on a glass substrate ($S_{i}O_{2}$) with the smallest line width of $ 0.152 \; \mu m$. It can be found that the \textit{500-}line pair pattern that cannot be clearly imaged by commercial Olympus objective (see Fig.~\ref{figure_4}(b)) can be clearly imaged through SODNN (see Fig.~\ref{figure_4}(c)), demonstrating that SODNN has performance comparable to the commercial microscopy imaging system. A more compact solution is to integrate SODNN with optical fiber to form an endoscope, as described in Discussion section 4.4.

\section*{4 Discussion}\label{disscussion}

\subsection*{4.1 Comparisons with State-of-the-art Methods}\label{4_1}

Table 1 shows several typical achievements of super-oscillatory design compared with SODNN. SODNN can achieve a long working distance with hundreds of micrometers, while the other works can only achieve a short working distance with only tens of micrometers.  In terms of working distance, SODNN has improved by an order of magnitude. SODNN can also achieve a DoF greater than 6 $\mu$m, and the ratio of FWHM to Rayleigh diffraction limit is less than 60\%, which demonstrates the extraordinary performance of SODNN.

\begin{figure*}[!ht]
\small Table 1 Comparisons of SODNNs with State-of-the-art Super-oscillatory Methods
    \centering
    \includegraphics[width=1.0\textwidth]{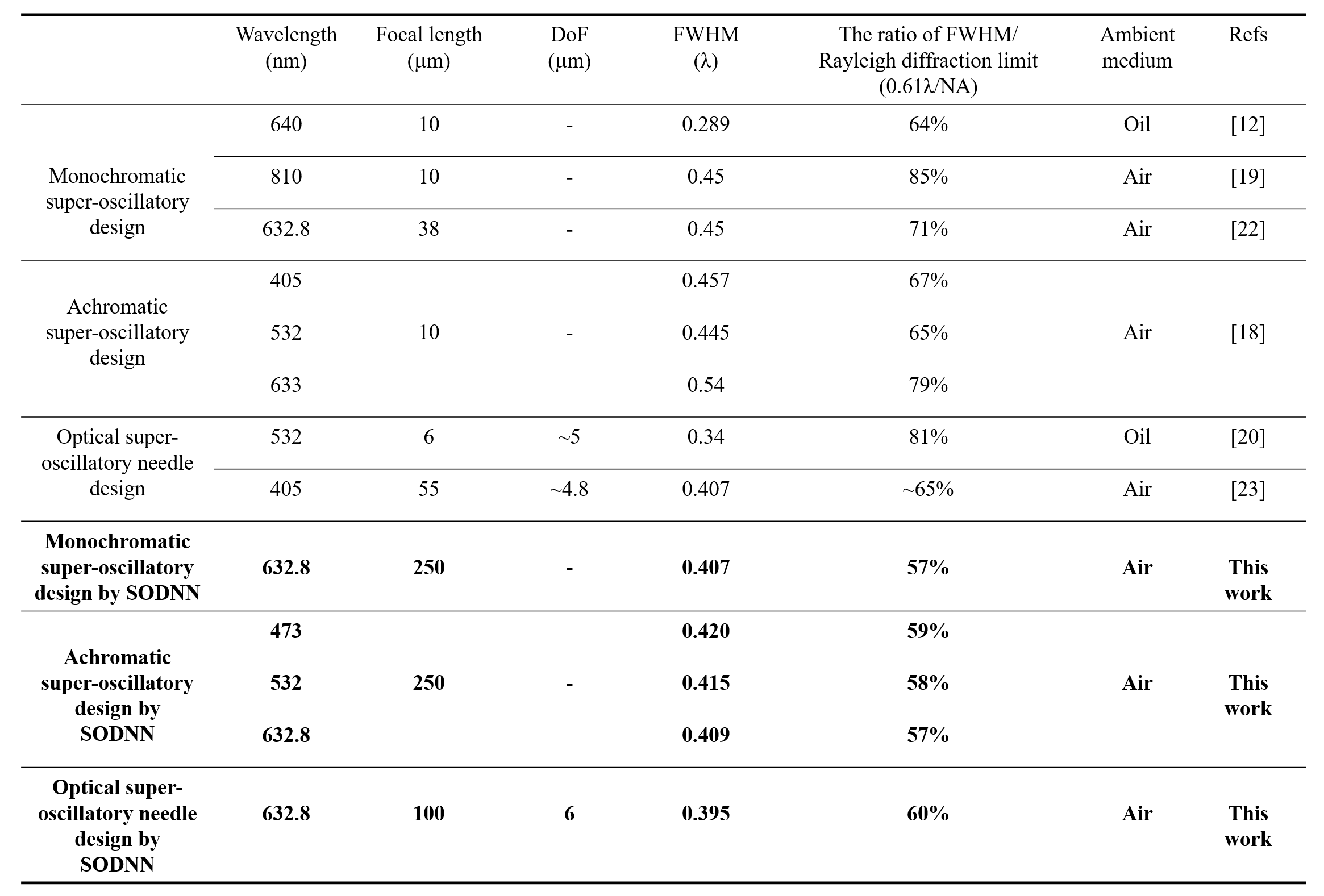}
    % \caption{\textbf{Performance analysis of SODNN as a function of the (a) modulation element numbers, (b) layer numbers and (c) diffractive element sizes.}}
    \label{figure_5}
\end{figure*}

\subsection*{4.2 Performance Analysis of SODNN}\label{4_2}
\begin{figure*}[!ht]
    \centering
    \includegraphics[width=1.0\textwidth]{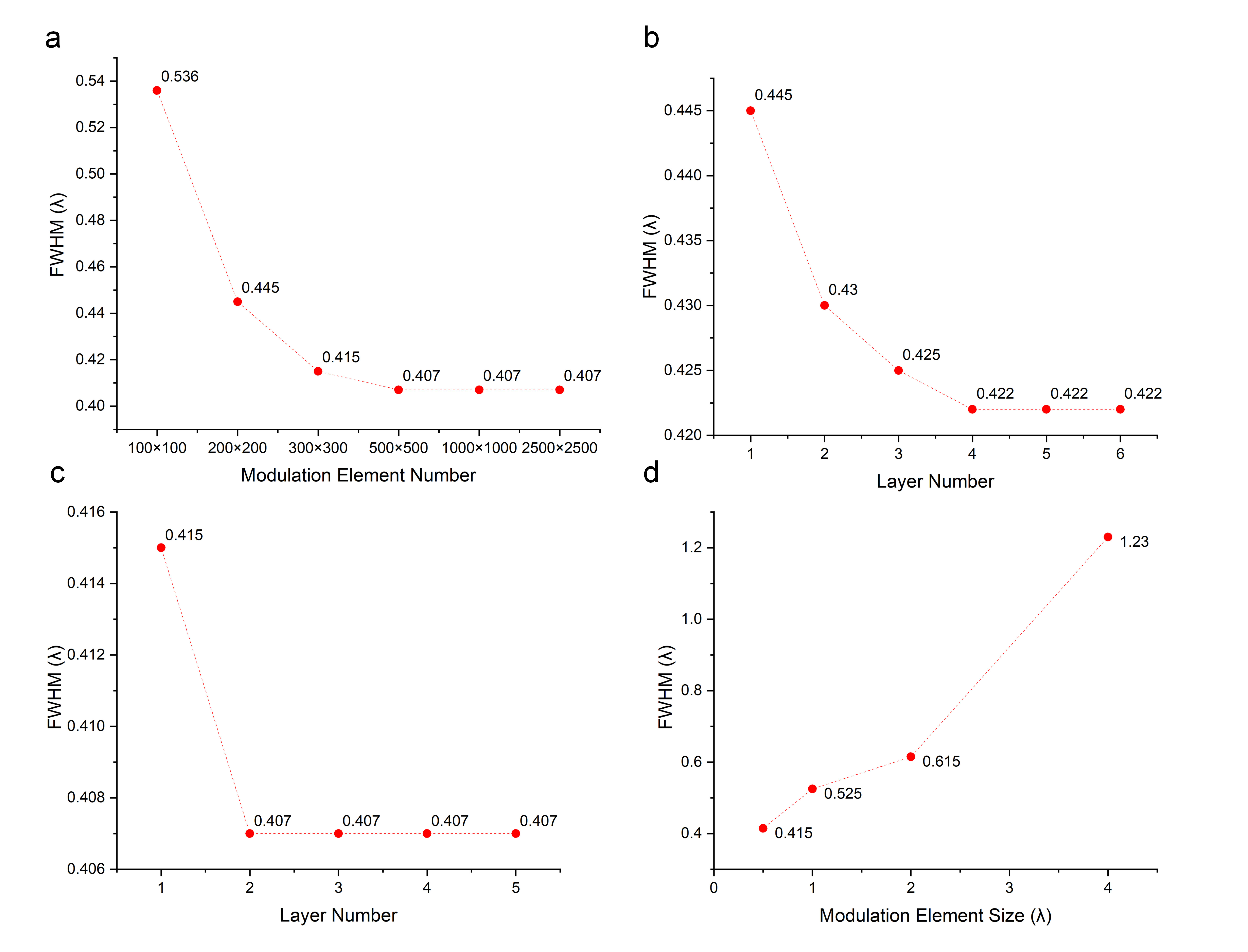}
    \caption{\textbf{Performance analysis of SODNN.} The FWHM of the output spot with respect to the modulation element number (a), layer number (b and c), and modulation element sizes (d).}
    \label{figure_6}
\end{figure*}
We evaluate and compare the network performance under different modulation element numbers, i.e., K × K, K = 100, 200, 300, 500, 1000, and 2500 (Fig.~\ref{figure_6}(a)) when the number of layers is fixed at 1, and diffractive element size is $\lambda$/2 × $\lambda$/2. We also evaluate the network performance under different layer numbers when the modulation element numbers are fixed at 200 × 200 (Fig.~\ref{figure_6}(b)) and 300 × 300 (Fig.~\ref{figure_6}(c)), and the diffractive element size is $\lambda$/2 × $\lambda$/2. We also analyze the effects of changes in neuron size, i.e., the diffractive element size was set to $\lambda$/2 × $\lambda$/2, $\lambda$ × $\lambda$, 2$\lambda$ × 2$\lambda$ and 4$\lambda$ × 4$\lambda$ when the number of layers is fixed at 1 and modulation element number is 300 × 300 (see Fig.~\ref{figure_6}(d)). We found that as the number of layers increases and the number of neurons in each layer increases, the FWHM of the super-oscillatory spot without side lobes will gradually become smaller and stabilize at $\sim0.407\lambda$ (Fig.~\ref{figure_6}(a) and Fig.~\ref{figure_6}(c)). It seems that the modulation element numbers have a greater impact than the layer numbers on the FWHM by comparing Fig.~\ref{figure_6}(b) and Fig.~\ref{figure_6}(c). When the modulation element numbers are small, no matter how the layer numbers increased, the FWHM of the super-oscillatory spot without side lobes cannot be stabilized at $\sim0.407\lambda$. On the contrary, as the diffractive element size increases, the FWHM of the super-oscillatory spot without side lobes will gradually increase. The above results will guide system design so that under specific system parameters, e.g., a 1-layer SODNN with enough modulation element numbers can achieve optimal performance as a multi-layer SODNN, thereby greatly reducing hardware complexity, system errors, and experimental difficulty.

\subsection*{4.3 Reconfigurable SODNN for Super-oscillatory Spot Scanning}\label{4_3}

\begin{figure*}[!h]
    \centering
    \includegraphics[width=0.8\textwidth]{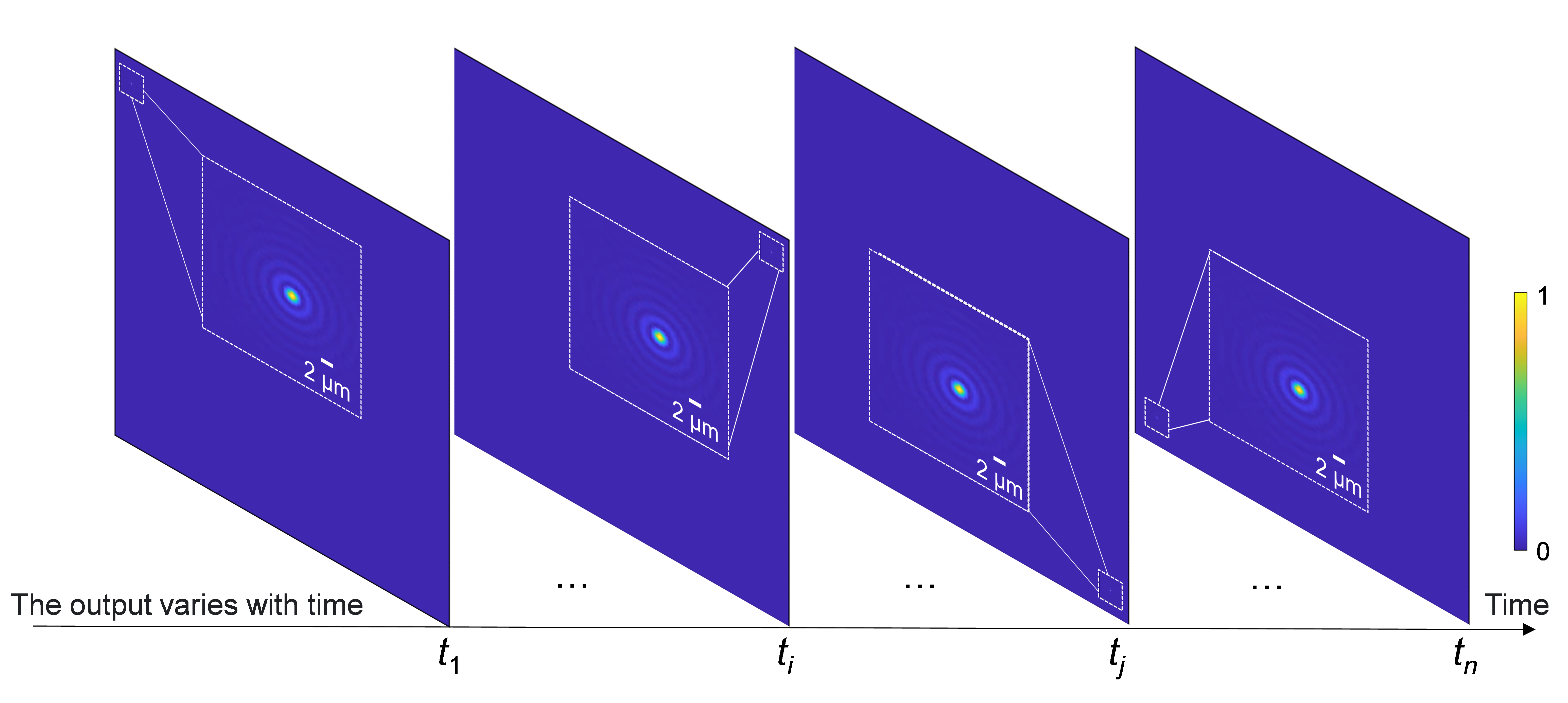}
    \caption{\textbf{Reconfigurable SODNN for super-oscillatory imaging.} The super-oscillatory spot can raster scan over the field for imaging by dynamically programming the modulation coefficients of diffractive elements.}
    \label{figure_7}
\end{figure*}

SODNN can modulate the incident optical field to create optical super-oscillatory effects in any 3D space and generate the super-resolved focusing spots. By training a series of SODNNs and loading them into a reconfigurable device such as a spatial light modulator (SLM), we can achieve dynamic scanning of super-oscillatory spots. We designed a series of SODNNs by setting the modulation element number to 2500 × 2500 with the element size $\lambda$/2 × $\lambda$/2 and $\lambda = 632.8 \; nm$, corresponding to the network layer size of ~0.79 $mm$ × 0.79 $mm$ with the focal length $ \emph{f}=250 \; \mu m $. The FWHM remains at  258 $nm$ ($ \sim0.407 \lambda $) and does not change with the position change of the super-oscillatory spots. Fig.~\ref{figure_7} shows that SODNN can focus the super-oscillatory spots at even the four most edge locations of the detection plane, which proves that SODNN can realize scanning at any position on the detection plane. Considering a high-speed SLM that works at ~1000 $fps$ and takes the FWHM as the scanning interval, the SODNN can achieve a scanning range of ~66.56 $\mu m ^2 /s$.

\subsection*{4.4 Integrating SODNN with Optical Fiber}\label{Needle}
\begin{figure*}[!h]
    \centering
    \includegraphics[width=0.8\textwidth]{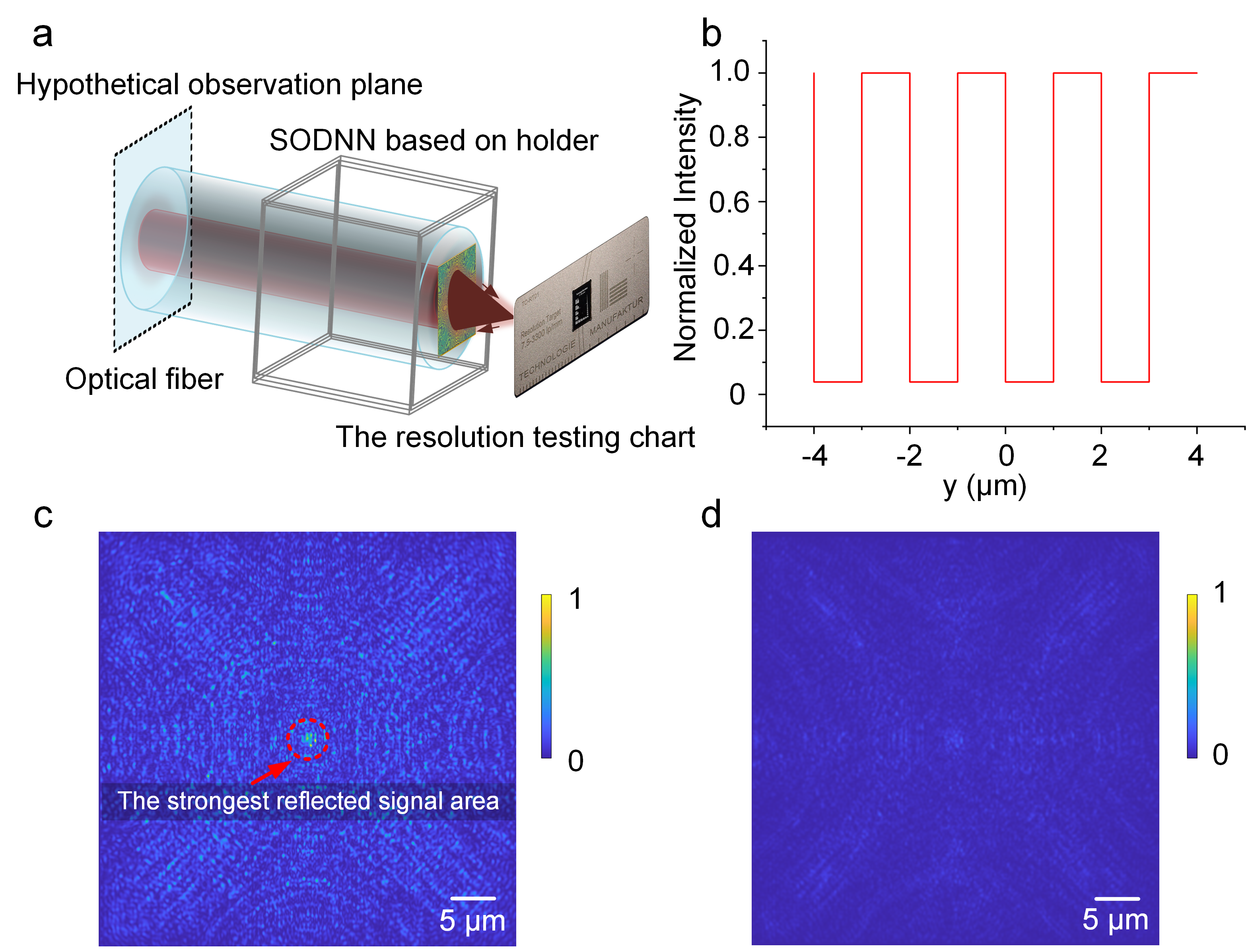}
    \caption{\textbf{Integrating SODNN with optical fiber.} (a) Schematic diagram of the endoscope designed by integrating SODNN and optical fiber. (b) The imaging function is realized by utilizing the intensity of reflected light produced by different transmittance structures, such as the metal structure producing a strong reflection (c) and the glass forming a weak reflection on the hypothetical observation plane (d).}
    \label{figure_8}
\end{figure*}
A more compact solution is to integrate SODNN with optical fiber to form an endoscope, as shown in Fig.~\ref{figure_8}(a), and collect the reflection signal of the super-oscillatory focusing spot from the measured target to complete the imaging process. Using Lumerical FDTD to arrange a hypothetical observation plane at the input end of SODNN to capture the topography of the reflected light field, we found the metal structure produces strong reflection (see Fig.~\ref{figure_8}(c)) with the normalized light intensity 1 and the glass forms a weak reflection with the normalized light intensity $\sim0$ (see Fig.~\ref{figure_8}(d)). Using the binary change of light intensity, we can also reconstruct the \textit{500-}line pair pattern of the measured resolution testing chart (see Fig.~\ref{figure_8}(b)).

\section*{5 Conclusion}\label{conclusion}
In conclusion, we proposed the SODNNs that demonstrate unparalleled advantages over other techniques for the realization of super-oscillatory spots and imaging beyond the diffraction limit. SODNN makes it possible to flexibly design large FoV without side-lobes, long working distances, long DoF, and achromatic optical super-oscillatory functions. SODNNs can work at any wavelength, from microwaves to ultraviolet waves \cite{r17}, which makes it possible to obtain super-oscillatory spots with smaller FWHM and further improve the resolution of SODNN imaging. We believe that this is a new cross-innovation in the fields of photonic neural networks, super-resolution microscopy, and metasurfaces, which will attract the attention of more scientists engaged in the development of intelligent optical instruments.

\bmhead*{Disclosures}
The authors declare no competing interests.

\bmhead*{Code, Data, and Materials Availability}
All relevant code is available from the corresponding author upon reasonable request.

\bmhead*{Acknowledgements}
This work is supported by the National Key Research and Development Program of China (grant 2021ZD0109902), the National Natural Science Foundation of China (grant 62275139), and the China Postdoctoral Science Foundation (grant 2023M741889).

\clearpage
\bibliographystyle{unsrt}

\clearpage 

\vspace{2ex}\noindent\textbf{Hang Chen} is a Postdoctoral Researcher and Research Associate at Tsinghua University. He received his BS degree, MS degree, and PhD degree in the School of Instrumentation Science and Engineering from Harbin Institute of Technology in 2015, 2017, and 2022, respectively. He is the author of more than 10 journal papers. His current research interests include photonic neural networks and integrated photonic chips. He is a young editorial board member of \textit{Acta Optica Sinica} and \textit{Chinese Laser Press}.

\vspace{2ex}\noindent\textbf{Sheng Gao} is a Ph.D. student in the Department of Electronic Engineering at Tsinghua University. He received his BE degree and MS degree in the School of Information and Electronics, Beijing Institute of Technology, in 2019 and 2022. His current research interests include photonic computing, diffractive neural networks, and electromagnetics.

\vspace{2ex}\noindent\textbf{Zejia Zhao} is a Ph.D. student in the Department of Electronic Engineering at Tsinghua University. She received her BE degree in the School of Optoelectronic Engineering at Xidian University in 2023. Her current research interests include optical nonlinearity, photonic computing, and metasurfaces.

\vspace{2ex}\noindent\textbf{Zhengyang Duan} is a Ph.D. student in the Department of Electronic Engineering at Tsinghua University. He received his BE degree in the Department of Electronic Engineering at Tsinghua University in 2022. His current research interests include the theory, design, and training of optical neural networks and optical-electronic heterogeneous computing architecture.

\vspace{2ex}\noindent\textbf{Haiou Zhang} is an Engineer in the Department of Electronic Engineering at Tsinghua University. She received her master's degree from Liaoning University in 2017. Her current research interests include photonic computing.

\vspace{2ex}\noindent\textbf{Gordon Wetzstein} is an Associate Professor of Electrical Engineering and, by courtesy, of Computer Science at Stanford University. He is the leader of the Stanford Computational Imaging Lab and a faculty co-director of the Stanford Center for Image Systems Engineering. At the intersection of computer graphics and vision, artificial intelligence, computational optics, and applied vision science, Prof. Wetzstein's research has a wide range of applications in next-generation imaging, wearable computing, and neural rendering systems. Prof. Wetzstein is a Fellow of Optica and the recipient of numerous awards, including an NSF CAREER Award, an Alfred P. Sloan Fellowship, an ACM SIGGRAPH Significant New Researcher Award, a Presidential Early Career Award for Scientists and Engineers (PECASE), an SPIE Early Career Achievement Award, an Electronic Imaging Scientist of the Year Award, an Alain Fournier Ph.D. Dissertation Award, as well as many Best Paper and Demo Awards.

\vspace{2ex}\noindent\textbf{Xing Lin} received a B.E. degree in Electronic Engineering from Xidian University, Xi’an, China, in 2010 and a Ph.D. degree in Automation from Tsinghua University, Beijing, China, in 2015. He was a Research Associate with Stanford University, Stanford, CA, USA, from 2015–2017, a Postdoctoral Scholar with the University of California, Los Angeles, Los Angeles, CA, USA, from 2019–2019, and a Research Scientist with Beijing-Tsinghua Innovation Center for Future Chips, Beijing, China, during 2019–2021. He is currently a tenure-track Assistant Professor with the Department of Electronic Engineering, Tsinghua University, Beijing, leading the Tsinghua Photonic Computing and Integration Research Lab. His research interests include photonic computing, neuromorphic photonics, and computational imaging. He was the recipient of MIT TR35 Asia Pacific, AI Chinese Young Scholar, and the Science and Technology Progress Award (first prize) of the Chinese Institute of Electronics.

\end{document}